Classification: Physical Sciences: Chemistry

The Unusual Superconducting State at 49 K in Electron-Doped $CaFe_2As_2$ at Ambient

B. Lv,[a] L. Z. Deng,[a,b] M. Gooch,[a] F. Y. Wei,[a,b] Y. Y. Sun,[a] J. Meen,[a,c] Y. Y. Xue,[a,b] B. Lorenz[a,b] and C. W. Chu[a,b,d]

[a]Texas Center for Superconductivity, University of Houston, Houston TX  77204-5002, USA

[b]Department of Physics, University of Houston, Houston TX  77204-5005, USA

[c]Department of Chemistry, University of Houston, Houston TX  77204-5003, USA

[d]Lawrence Berkeley National Laboratory, 1 Cyclotron Road, Berkeley CA  94720, USA

Corresponding Author:

C. W. Chu

Texas Center for Superconductivity

University of Houston

202 Houston Science Center

Houston TX  77204-5002

PHONE:   713-743-8222

FAX:   713-743-8201

email:   cwchu@uh.edu

**ABSTRACT**

We report the detection of unusual superconductivity up to 49 K in single crystalline $CaFe_2As_2$ via electron-doping by partial replacement of Ca by rare-earth. The superconducting transition observed suggests the possible existence of two phases: one starting at ~ 49 K, which has a low critical field ~ 4 Oe, and the other at ~ 21 K, with a much higher critical field > 5 T. Our observations are in strong contrast to previous reports of doping or pressurizing layered compounds $AeFe_2As_2$ (or Ae122), where Ae = Ca, Sr or Ba. In Ae122, hole-doping has been previously observed to generate superconductivity with a transition temperature ($T_c$) only up to 38 K and pressurization has been reported to produce superconductivity with a $T_c$ up to 30 K. The unusual 49 K phase detected will be discussed.

**Introduction**

The discovery of the layered Fe-pnictide 26 K superconductor LaFeAs(O,F) in 2008 has generated great excitement and hope in the high temperature superconductivity (HTS) community [1]. In the ensuing three years, extensive studies have been carried out worldwide in an attempt: 1) to unravel the role of magnetism in the occurrence of HTS because of the presence of a large amount of the magnetic Fe in the compounds, and 2) to explore the possibility of raising the $T_c$ to a higher value due to the existence of a large number of compounds isostructural to the layered Fe-pnictide superconductors. As a result, many layered Fe-pnictide superconductors have been found, although with a $T_c$ < 60 K as suggested [2]. They can be categorized into three phases, i.e. the 1111-phase (RFeAsO, where R = rare earth) [1] with the ZrCuSiAs structure (P4/nmm); the 122-phase ($AeFe_2As_2$ and $AFe_2As_2$, where Ae = alkaline earth and A = alkaline) [3] with the $ThCr_2Si_2$ structure (I4/mmm); and the 111-phase (AFeAs, where A = alkaline) [4] with the PbFCl structure (P4/nmm). All phases possess the tetrahedrally coordinated corner sharing FeAs layers, mainly through which the superconducting carriers flow. Fe-chalcogenide superconductors were also discovered, but with only two phase types loaded with vacancies, i.e. the 122-phase ($A_{1-x}Fe_{2-y}Se_2$ where A = alkaline) [5]; and the 11-phase ($FeSe_{1-x}$) [6] with the PbO structure (P4/nmm). They consist of the similar FeSe-layers to carry the majority of the conducting carriers. However, the maximum $T_c$s obtained in different families are 57 K for 1111 [7]; 38 K for 122 [3,8]; 18 - 31 K for 111 [4]; and 8 - 37 K for 11 [6], through doping and/or pressurization. In spite of the similar layered structures, $T_c$s of Fe-pnictides and -chalcogenides are much lower than those of the cuprates, which possess the $CuO_2$-layers [9]. While the $T_c$ of the cuprates increases with the complexity of the structures, such as the number of $CuO_2$-layers per unit cell, efforts to increase structure complexity have not yet yielded similar results in Fe-based superconductors.

It is known that chemical doping is possible only when the ionic-size-matching and charge-neutrality requirements are met. Until now, only electron-doping in the 1111 phase and hole-doping in the 122 phase have been reported to be successful at ambient pressure [10], although sporadic reports of doping opposite to the above have appeared. A few are still unsettled [11] and one was done under high pressure [12]. The existence of a symmetry between electron- and hole-doping has been demonstrated in cuprate HTSs [13] with respect to the induction of superconductivity in their respective parent compounds, although the situation in Fe-pnictides is still not clear [12]. However, the higher $T_c$ always occurs in the hole-doped cuprates, consistent with a model calculation [14]. On the other hand, for Fe-pnictides, the highest $T_c$ of 57 K occurs in the electron-doped 1111-phase. A question arises whether electron-doping can raise the $T_c$ of the 122-phase to above the current record of 38 K.

Due to the ionic radius matching constraint, we have electron-doped only Ca122 at ambient by partial replacement of Ca by Ce, Pr and Nd. All electron-doped Ca122 attempted became superconducting. However, the Pr-doped samples have the highest $T_c$ and the sharpest transition, perhaps due to the better ionic matching between $Ca^{2+}$ and $Pr^{3+}$. Here we report the attainment of superconductivity with an onset $T_c$ of ~ 49 K in Pr-doped Ca122 as evidenced from the resistive, magnetic and thermoelectric measurements. The superconducting transition exhibits an unusual magnetic field dependence, suggestive of the existence of two superconducting phases, one starting at ~ 49 K and the other at ~ 21 K. A first-order structural phase transformation is also detected slightly above $T_c$. The results will be presented and their implications discussed.

**Experimentals**

Single crystals of electron-doped Ca122 with Pr were successfully grown from self-flux. The FeAs precursor was first synthesized from stoichiometric amounts of Fe (99.999+% from Aldrich) and As (99.9999% from Alfa) inside the silica tube at 800 °C for 30 hours. Then Pr-pieces (99.9% from Alfa) and Ca-pieces (99.99% from Alfa) were mixed with FeAs according to the ratio of (Pr+Ca)/FeAs = 1/4 and placed in an alumina crucible inside a silica tube sealed under reduced Ar atmosphere. The silica tube was subsequently sealed inside a larger silica tube under vacuum to prevent the sample from getting into contact with air if the first tube failed. The assembly was then put inside a box furnace, heated to 1200 °C for 8 hours, and then cooled to 980 °C slowly at 2 °C/hr. The sample was finally furnace-cooled to room temperature by turning off the power. Single crystals with the flat shiny surface up to 5 mm x 5 mm size were easily cleaved from the melt. All the preparative manipulations were carried out in a purified argon atmosphere glove box with a total $O_2$ and $H_2O$ level < 1 ppm.

Crystalline samples were characterized by X-ray diffraction using a Rigaku DMAX III-B diffractometer. Chemical analyses were performed using wavelength-dispersive spectrometry (WDS) on a JEOL JXA-8600 electron microprobe analyzer with 15 kV accelerating voltage, a 30 nA sample current, and 1 μm spot size. Precision of the results is smaller than 0.5% relative, and quoted errors reflect variations of count rates in multiple analyses of samples and exceed the precision of each individual analysis.

Electrical resistivity as a function of temperature ρ(T) was measured by employing a standard 4-probe method using a Linear Research LR-400 ac bridge operated at 15.9 Hz. The temperature dependence of the dc- and ac-magnetic susceptibility, $\chi_{dc}$ and $\chi_{ac}$(T), was measured using a Quantum Design Magnetometer with the superconducting quantum interference device (SQUID) at fields up to 5 T. Thermoelectric power S(T) was measured using a low frequency (0.1 Hz) ac technique with a resolution of 0.02 μV/K. During the measurements, the amplitude of the sinusoidal temperature modulation was kept constant at 0.25 K. The specific heat $C_p$(T) was determined using the Quantum Design PPMS system.

**Results and Discussion**

Typical XRD pattern of the single crystalline samples exhibits the expected preferred orientation along the c-axis, as shown in Fig. 1. However, the relative line-intensities do not compare well with the simulation and the theta scan of the *008*-peak shows a spread that increases with Pr-doping up to 2.5°, indicating imperfection in crystallinity of the sample induced by doping. Detailed structural analyses will be published elsewhere. Single crystalline samples of $Ca_{1-x}Pr_xFe_2As_2$ with nominal x = 0.05 – 0.24 prepared according to the above mentioned conditions have the real $x_{WDS}$ of 0.059 – 0.127. All show zero-ρ above 15 K. However, samples with $x_{WDS}$ of 0.121 – 0.127 show sharp and narrow resistive superconducting transitions with a resistive onset-$T_c$ as high as ~ 49 K, as exemplified by Sample #263 with $x_{WDS}$ = 0.127 in the inset of Fig. 2. ρ(T) of this sample exhibits a small tail possibly due to crystal imperfection, and a curvature change with temperature typical of a strongly correlated electron system as displayed in Fig. 2. The effect of magnetic field on ρ(T) of Sample #264 with $x_{WDS}$ = 0.121 is shown in Fig. 3. Little ρ(T)-change is detected at fields below 200 Oe but the superconducting transition is broadened and shifted to lower temperature above. Superconductivity survives at 5 T, the maximum field of our experiment. The thermoelectric power S(T) shown in Fig. 4 displays three features on cooling: it is negative at room temperature, suggesting that the carriers are mainly electron in nature as expected; shows a hysteretic anomaly around ~ 70 K, characteristic of a first order phase transition; and rises rapidly at ~ 49 K and becomes zero at ~ 44 K, indicative of the entering into a superconducting state. The zero-field-cooled dc magnetic susceptibility zfc $\chi_{dc}$(T) of Sample #263 is displayed in Fig. 5 at different fields. At 1 Oe, a clear diamagnetic signal characteristic of a superconducting transition with an estimated onset-$T_c$ ~ 47 K and a maximum superconducting shielding signal corresponding to ~ 40% at 5 K after demagnetization correction (demagnetization factor of ~ 7) are obtained. The field-cooled $\chi_{dc}$(T) shows a large trapped field below ~ 47 K and an additional field expulsion below ~ 21 K, coinciding with a rapid drop in the zfc $\chi_{dc}$(T), as exemplified in the inset of Fig. 5 at 1 Oe. As the field increases, both the onset-$T_c$ and the superconducting signal decrease rapidly In fact, the negative $\chi_{dc}$(T) of superconductivity changes sign at 300 Oe and becomes positive above. This is very different from the field effect on ρ(T) shown in Fig. 3, where a field of 5 T is not sufficient to destroy superconductivity. The magnetization vs. field (M-H) loop was determined and is shown in Fig. 6, with the low field data given in the insets, which gives a $H_{c1}$ < 4 Oe and a $H_{c2}$ > 5 T at 5 K. In addition, a paramagnetic background is in evidence. To reconcile the apparent conflicting field effects on ρ(T) and $\chi_{dc}$(T), we reduce the possible interference from the magnetic background of the sample with the superconducting signal and measure the ac magnetic susceptibility $\chi_{ac}$(T) under different dc fields. The results are shown in Figs. 7 and 8. The $\chi_{ac}$(T) at 0 Oe in Fig. 7 clearly exhibits two parts of the transition, with drastically different responses to the magnetic field, i.e. the high temperature part with an onset-$T_c$ ~ 47 K is suppressed quickly by a field ~ 300 Oe down to ~ 21 K as displayed in Fig. 8 and the low temperature part with an onset-$T_c$ ~ 21 K is only moderately suppressed by a field up to 5 T as shown in Fig. 7, suggesting the possible existence of two superconducting phases, consistent with the ρ(T) and $\chi_{dc}$(T)-data in Figs. 3 and 5, respectively. The $C_p$(T) after subtraction of the phonon background between 5 and 160 K shows a clear anomaly at ~ 60 K but fails to reveal unambiguously the anomaly associated with a superconducting transition. The ~ 60 K anomaly is consistent with the S(T)-anomaly observed in Fig. 4, which is attributed to a first-order structural transition.

The above observations have firmly established that Pr-doping induces superconductivity in Ca122 up to 49 K, higher than any $T_c$ previously reported in the 122-phase. The high $T_c$ detected in Pr-doped Ca122 is puzzling, in view of the structural and chemical similarities of Ca, Sr and Ba122 and the same maximum $T_c$ of the doping-induced (~ 38 K) or pressure-induced (~ 30 K) superconductivity in Sr and Ba122. This is particularly so since the maximum $T_c$ of the doping- or pressure-induced superconductivity in Ca122 (~ 26 K or ~ 10 K) is always lower than that in Sr122 and Ba122 and even the nature of the pressure-induced superconducting state in Ca122, i.e. filamentary or bulk, remains unsettled [15].

The different responses of $\chi_{ac}(T,H)$ to fields in different temperature ranges shown in Figs. 7 and 8, i.e. below 21 K and above, strongly suggest that there are two superconducting transitions in the sample: one with an onset-$T_c$ ~ 49 K and the other, ~ 21 K. The diamagnetic $\chi_{ac}(T)$ of superconductivity above 21 K is almost totally suppressed by a field > 500 Oe, while that below 21 K is less field-dependent. Even at 5 T, the highest field applied in this experiment, superconductivity remains below ~ 15 K. Since the overall superconducting transition as indicated by $\chi_{dc}(T)$ and $\chi_{ac}(T)$ shown in Figs. 5 and 7 has not completed down to 5 K, the decomposition of the transition into two, with one representing the high $T_c$ phase and the other, the low $T_c$ phase, can only be taken qualitatively. The high $T_c$-phase above ~ 21 K can be represented by $\chi_{ach}(T>21K,H)$ = $\chi_{ac}(T>21K,H)$ - $\chi_{ac}(T>21K,500\ Oe) \approx \chi_{ac}(T>21K,H)$ since the high $T_c$-phase is completely suppressed by 500 Oe, and is shown in Fig. 8 for H < 500 Oe. Similarly, the low $T_c$-phase can be represented by $\chi_{acl}(T<21K,H) = \chi_{ac}(T<21K,H) - \chi_{ach}(T<21K,H)$ $\approx \chi_{ac}(T<21K,H)$ since the high $T_c$-phase is completely suppressed, and is displayed in Fig. 7. One can easily see that the field effects on the two superconducting phases are totally different.

The two transitions appear to be readily understood by assuming that they are due to the grains usually present in a ceramic polycrystalline sample. The high temperature transition is then caused by the Josephson junction coupling across the grains and the low temperature transition realized when the coupling between grains becomes stronger on cooling when a phase-lock transition between grains is achieved. While such a picture accounts for the observation in granular polycrystalline superconductors well, it does not address the single crystalline materials associated with the two transitions with two different $T_c$s. There are three possible scenarios: 1) one phase of Pr-doped Ca122 grains throughout the sample; 2) one phase of Pr-doped Ca122 with part of the grain misaligned; or 3) a minor unknown phase, perhaps due to chemical inhomogeneity, embedded in the matrix of Pr-doped Ca122. The XRD results in Fig. 1 show that the relatively good crystallinity of the sample is not consistent with scenario #1 but is compatible with scenarios #2 and #3. The less-than-perfect XRD pattern as described earlier provides room for the small misalignment of the small grains present in the samples and/or minor phase in the grain boundary. Although the 21 K transition may be attributed to the bulk Pr-doped Ca122, the 49 K transition cannot, since no $T_c$ > 38 K has been reported in the equilibrium Pr-Ca-Fe-As compound system. Although electron-doping in Ca122 and possible soft phonons associated with the first order structural transition near and above $T_c$ may be beneficial to higher $T_c$, given the granular and filamentary nature of the 49 K transition, we conjecture that the high $T_c$ may be associated with a metastable phase or a phase of special connectivity of Pr-Ca-Fe-As. The high pressure work on Ca122 has demonstrated that a strain-induced metastable filamentary or inter-gain superconducting phase can exist in Ca122 under pressure [15]. It has also been shown that interfacial or filamentary superconductivity can have an enhanced $T_c$ [16]. The present experiment may have shown such a possibility, although possibly chemically induced. Given the potential significance of the discovery of interfacial and/or filamentary superconductivity with enhanced $T_c$, further work is warranted to confirm or refute this proposition. It should be also noted that the very sensitive dependence of the high $T_c$-phase on field suggests the transition at ~ 49 K may be of Josephson-Junction coupling between grains in nature. This may imply the existence of a superconducting phase with a $T_c$ higher than 49 K in the compound system investigated here.

## Note

At the completion of our work, we learned that a similar $T_c$ in Pr-doped Ca122 was also observed by Saha et al. (arXiv:1105.4798v1 [cond-mat.supr-con]), although with an emphasis on the collapsed phase, at the U.S. Air Force Office of Scientific Research Joint Electronics Program Review in Arlington, Virginia, on May 23, 2011.

## Acknowledgement

The work in Houston is supported in part by AFOSR No. FA9550-09-1-0656, DoE subcontract 4000086706 through ORNL, AFRL subcontract R15901 (CONTACT) through Rice University, the T. L. L. Temple Foundation and TCSUH; and at LBNL by the Director, Office of Science, OBES, DMSE, DoE.

## References

1. Y. Kamihara Y, Watanabe T, Hirano M, Hosono H (2008) Iron-based layered superconductor La[O$_{1-x}$F$_x$]FeAs (x = 0.05-0.12) with $T_c$ = 26 K. *J Am Chem Soc* 130:3296-3297.
2. Lorenz B, et al. (2008) Effect of pressure on the superconducting and spin-density-wave states of SmFeAsO$_{1-x}$F$_x$. Phys Rev B 78:012505.
3. Rotter M, Tegel M, Johrendt D (2008) Superconductivity at 38 K in the iron arsenide (Ba$_{1-x}$K$_x$)Fe$_2$As$_2$. *Phys Rev Lett* 101:107006; Sasmal K, et al. (2008) Superconducting Fe-based compounds (A$_{1-x}$Sr$_x$)Fe$_2$As$_2$ with A = K and Cs with transition temperatures up to 37 K. *Phys Rev Lett* 101:107007.
4. Tapp JH, et al. (2008) LiFeAs: An intrinsic FeAs-based superconductor with $T_c$ = 18 K. *Phys Rev B* 78:060505(R); Zhang SJ, et al. (2009) Superconductivity at 31 K in the "111"-type iron arsenide superconductor Na$_{1-x}$FeAs induced by pressure. *Europhys Lett* 88:47008.
5. Guo JG, et al. (2010) Superconductivity in the iron selenide K$_x$Fe$_2$Se$_2$ (0 ≤ x ≤ 1.0). *Phys Rev B* 82:180520(R).

6. Hsu FC, et al. (2008) Superconductivity in the PbO-type structure α-FeSe. *Proc Natl Acad Sci USA* 105:14262-14264; Margadonna S, et al. (2009) Pressure evolution of the low-temperature crystal structure and bonding of the superconductor FeSe ($T_c$ = 37 K). *Phys Rev B* 80:064506.
7. Chen XH, et al. (2008) Superconductivity at 43 K in $SmFeAsO_{1-x}F_x$. *Nature* 453:761-762; Ren ZA, et al. (2008) Superconductivity in the iron-based F-doped layered quaternary compound $Nd[O_{1-x}F_x]FeAs$. *Europhys Lett* 82:57002.
8. Sasmal K, et al. (2008) Superconducting Fe-based compounds $(A_{1-x}Sr_x)Fe_2As_2$ with A = K and Cs with transition temperatures up to 37 K. *Phys Rev Lett* 101:107007; Gooch M, Lv B, Lorenz B, Guloy AM, Chu CW (2008) Pressure-induced shift of $T_c$ in $K_xSr_{1-x}Fe_2As_2$ (x = 0.2,0.4,0.7): Analogy to the high-$T_c$ cuprate superconductors. *Phys Rev B* 78:180508(R).
9. Chu CW (2011) in *100 Years of Superconductivity*, eds Rogalla H, Kes PH (Chapman & Hall/CRC), Chapter 4 (in press).
10. Ishida K, Nakai Y, Hosono H (2009) To what extent iron-pnictide new superconductors have been clarified: A progress report. *J Phys Soc Jpn* 78:062001; Johnston DC (2010) The puzzle of high temperature superconductivity in layered iron pnictides and chalcogenides. *Advances in Physics* 59:803-1061; Paglione J and Greene RL (2010) High-temperature superconductivity in iron-based materials. *Nature Physics* 29:645-658.
11. Wen HH, Mu G, Fang L, Yang H, Zhu ZY (2008) Superconductivity at 25 K in hole-doped $(La_{1-x}Sr_x)OFeAs$. *Europhys Lett* 82:17009; Wu G, et al. (2008) Superconductivity induced by oxygen deficiency in $La_{0.85}Sr_{0.15}FeAsO_{1-\delta}$. *Phys Rev B* 78:092503.
12. Muraba Y, et al. (2010) High-pressure synthesis of the indirectly electron-doped iron pnictide superconductor $Sr_{1-x}La_xFe_2As_2$ with maximum $T_c$ = 22 K. *Phys Rev B* 82:180512(R).
13. Tokura Y, Takagi H, Uchida S (1989) A superconducting copper oxide compound with electrons as the charge carriers. Nature 337:345-347; Mazin II (2010) Superconductivity gets an iron boost. Nature 464:183-186.
14. Marsiglio F and Hirsch JE (2008) Hole superconductivity in arsenic-iron compounds. *Physica C* 468:1047-1052.

15. Goldman AI, et al. (2009) Lattice collapse and quenching of magnetism in $CaFe_2As_2$ under pressure: A single-crystal neutron and x-ray diffraction investigation. *Phys Rev B* 79:024513; Yu W, et al. (2009) Absence of superconductivity in single-phase $CaFe_2As_2$ under hydrostatic pressure. *Phys Rev B* 79:020511(R).
16. Allender D, Bray J, Bardeen, J (1973) Model for an exciton mechanism of superconductivity. *Phys Rev B* 7:1020-1029.

**Figure Captions**

Fig. 1   XRD of the Pr-doped Ca122.

Fig. 2   The $\rho(T)$ of sample #263. Inset: the low temperature $\rho(T)$.

Fig. 3   The magnetic field effect on $\rho(T)$ (Sample # 264).

Fig. 4   The S(T). Inset: the low temperature warming and cooling (Sample #264).

Fig. 5   The zfc $\chi_{dc}(T)$ after demagnetization correction at different fields: 1—1 Oe; 2—5 Oe; 3—10 Oe; 4—20 Oe; 5—100 Oe; 6—300 Oe (Sample #263). Inset: the zfc $\chi_{dc}(T)$ and fc $\chi_{dc}(T)$.

Fig. 6   The magnetization (M) vs. field (H) at 5 K. Lower inset: M-H hysteresis loop at fields below 100 Oe. Upper inset: Low-field data revealing the small lower critical field, $H_{c1}$ (Sample #263).

Fig. 7   The $\chi_{ac}(T)$ at different fields up to 5 T.

Fig. 8   The $\chi_{ac}(T)$ at different fields lower than 0.01 T.

Fig. 1

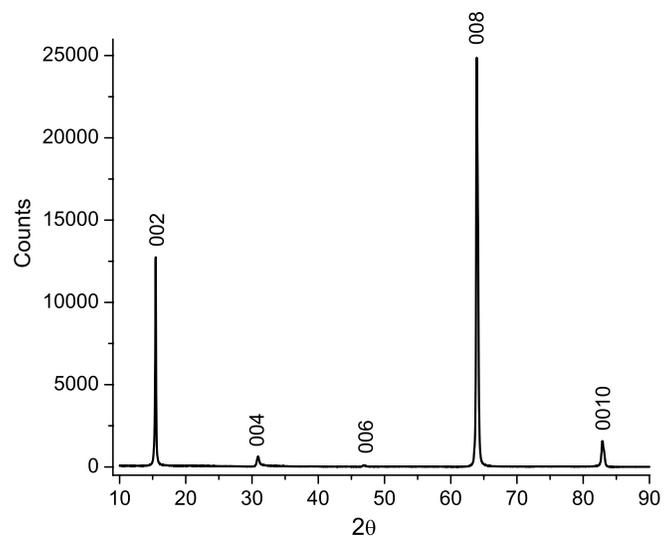

Fig. 2

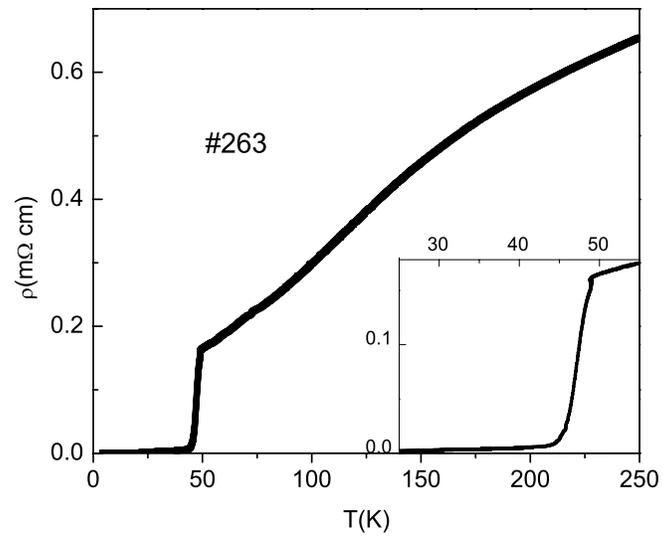

Fig. 3

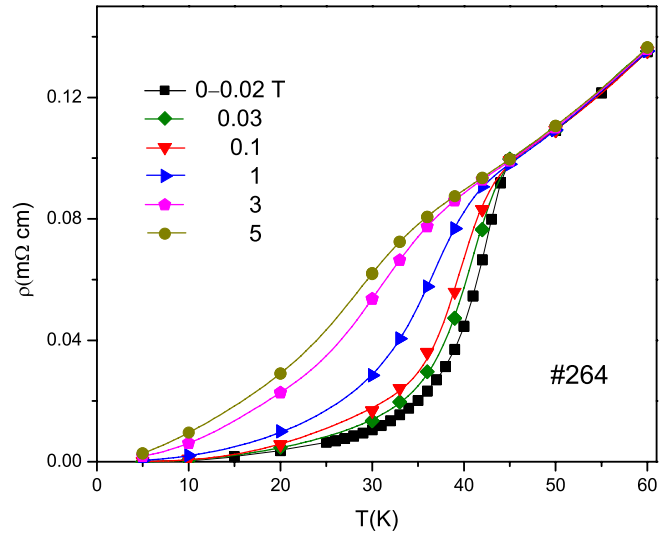

Fig. 4

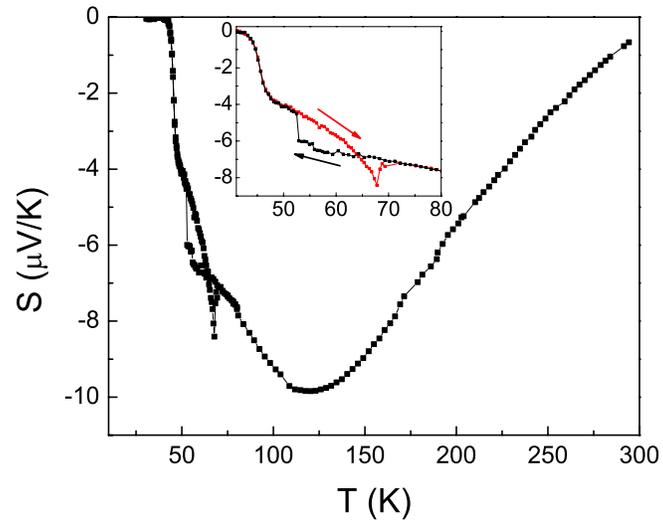

Fig. 5

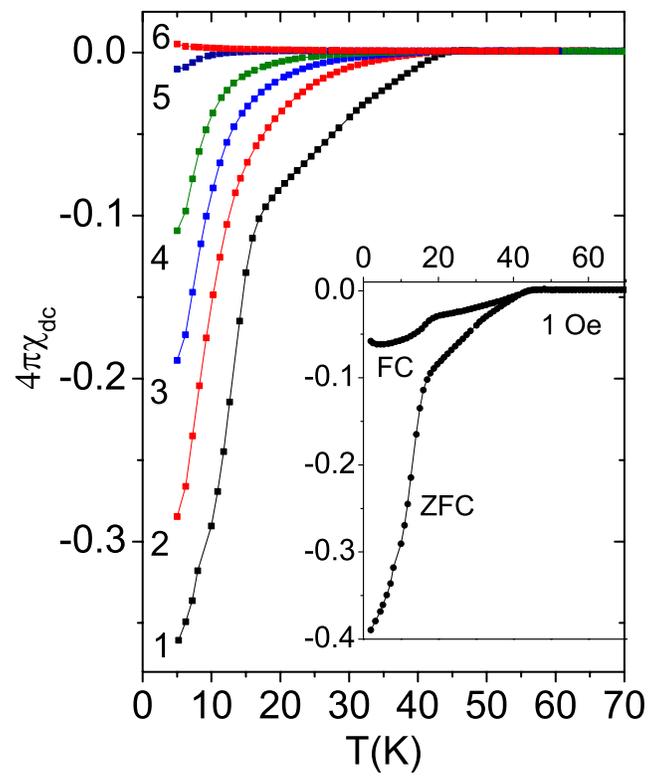

Fig. 6

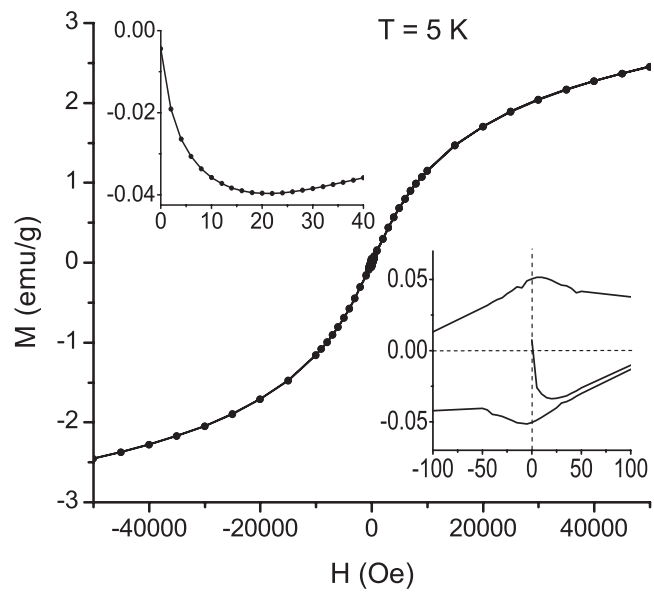

Fig. 7

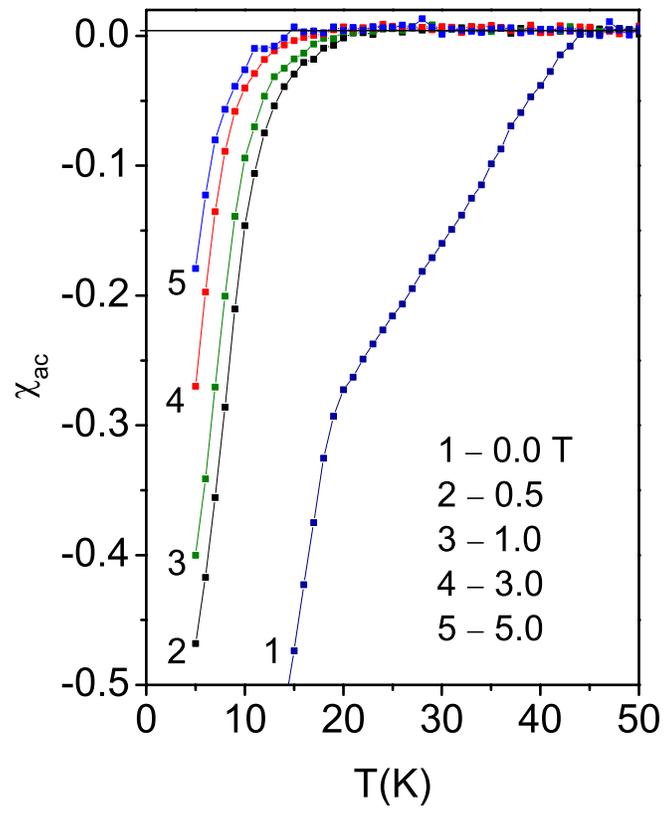

Fig. 8

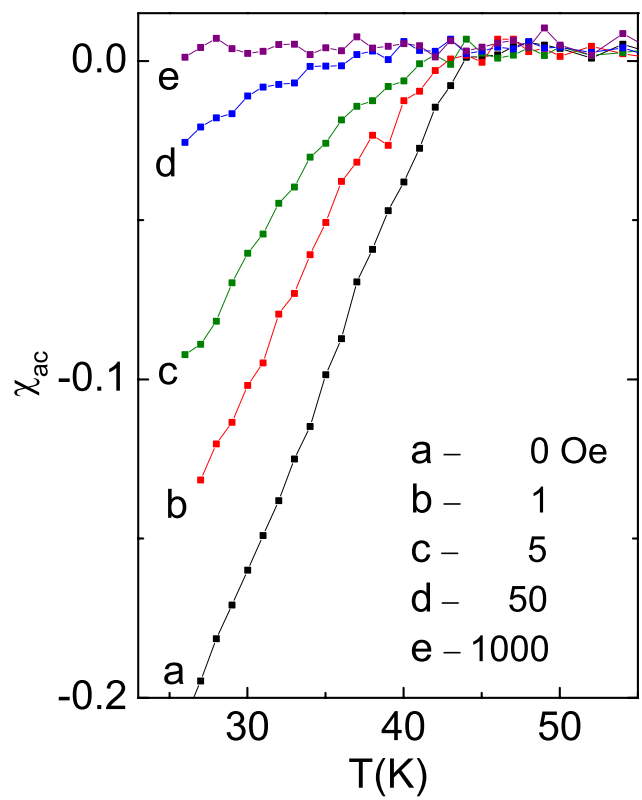